\documentclass[twocolumn]{autart}

\usepackage{graphicx}          
\usepackage[dvips]{epsfig}    

\usepackage{dblfloatfix}
\usepackage{fixltx2e}
\usepackage{dsfont} 

\def\c#1{{\rm{#1}}}
\def\ee{{\rm e}}
\newcommand {\bsis} {\left\{ \begin{array} }
\newcommand {\esis} {\end{array}\right.}
\def\Sum#1#2{\sum\limits_{#1}^{#2}}
\def\fracg#1#2{{\displaystyle{\frac{#1}{#2}}}}  
\def\set#1#2{\{ \; #1 \;:\;#2\;\}} 
\def\conv#1#2{ \left( \begin{array}{c} #1 \\ #2 \\
\end{array}\right)} 

\newcommand {\bmat} {\left[\begin{array} }

\newcommand {\emat} {\end{array}\right]}

\newcommand{\blista}{\renewcommand{\labelenumi}{(\roman{enumi})} 
\begin{enumerate}}
\newcommand{\elista}{\end{enumerate} \renewcommand{\labelenumi}{\arabic{enumi}.}}

\newcommand{\ourproof}{{\bf Proof: }}
\newcommand{\nt}{{n_{\theta}}} 
\newcommand{\pV}{E}

\newcommand{\setW}{{\mathcal{W}}}
\newcommand{\setD}{{\mathcal{D}}}
\newcommand{\setV}{{\mathcal{V}}}

\newcommand{\nameE}{{probability of violation}}

\newcommand{\namep}{probability of failure}

\newtheorem {assumption}{Assumption}
\newtheorem {definition}{Definition}
\newtheorem{lemma}{Lemma}
\newtheorem {property}{Property}
\newtheorem {remark}{Remark}
\newtheorem{theorem}{Theorem}
\newtheorem{corollary}{Corollary}

\newtheorem {example}{ Example}

\begin{document}

\begin{frontmatter}

\title{Randomized Methods for Design of Uncertain Systems:
Sample Complexity and Sequential Algorithms}

\author[US]{T. Alamo}\ead{alamo@cartuja.us.es},
\author[tempo]{R. Tempo}\ead{roberto.tempo@polito.it},
\author[US]{A. Luque}\ead{amalia@cartuja.us.es},
\author[US]{D.R. Ramirez}\ead{danirr@cartuja.us.es}

\address[US]{Departamento de Ingenier\'{\i}a de Sistemas y Autom\'{a}tica, Universidad de Sevilla,
Escuela Superior de Ingenieros, Camino de los Descubrimientos s/n,
41092 Sevilla. Spain}  
\address[tempo]{CNR-IEIIT, Politecnico di Torino, Corso Duca
degli Abruzzi 24, Torino 10129, Italy}             


\begin{abstract}                          

In this paper, we study randomized methods for feedback design of uncertain systems. The first contribution is to derive the sample complexity of various constrained control problems.
In particular, we show the key role played by the binomial distribution and related tail inequalities, and compute the sample complexity. This contribution significantly improves the existing results by reducing the number of required samples in the randomized algorithm. These results are then applied to the analysis of worst-case performance and design with robust optimization.
The second contribution of the paper is to introduce a general class of sequential algorithms, denoted as Sequential Probabilistic Validation (SPV). In these sequential algorithms, at each iteration, a candidate solution is probabilistically validated, and corrected if necessary, to meet the required specifications.
The results we derive provide the sample complexity which
guarantees that the solutions obtained with SPV algorithms meet some pre-specified
probabilistic accuracy and confidence. The performance of these algorithms is illustrated and compared with other existing methods using a numerical example dealing with robust system identification.

\end{abstract}

\end{frontmatter}
\section{Introduction}\label{sec:randomized:approach}
The use of randomized algorithms for systems and control has matured thanks to the considerable research efforts made in recent years. Key areas where we have seen convincing developments include uncertain and hybrid systems \cite{Tempo13,Vidyasagar97}. A salient feature of this approach is the use of the theory of rare events and large deviation inequalities, which suitably bound the tail of the probability distribution. These inequalities are crucial in the area of statistical learning theory \cite{Vapnik98}, which has been utilized for feedback design of uncertain systems \cite{Vidyasagar01a}.

Design in the presence of uncertainty is of major relevance in different areas, including
mathematical optimization and robustness \cite{BenNem:98,PetTem:14}.
The goal is to find a feasible solution which is optimal in some sense for all possible uncertainty instances.
Unfortunately, the related semi-infinite optimization problems are often NP-hard (examples of NP-hard problems in systems and control can be found in   \cite{Blondel97,Blondel00Aut}), and this may seriously limit their applicability from the computational point of view. There are two approaches to resolve this NP-hard issue. The first approach is based on the computation of deterministic relaxations of the original problem, which are usually polynomial time solvable. However, this might lead to overly conservative solutions \cite{Sherer06}. An
alternative is to assume that a probabilistic description of the
uncertainty is available. Then, a randomized algorithm may be
developed to compute, in polynomial time, a solution with
probabilistic guarantees \cite{Tempo13,Vidyasagar97}.
Stochastic programming methods \cite{Prekopa:95} are similar in spirit to the methods studied in this paper and take advantage that, for random uncertainty,  the underlying probability distributions are known or can be estimated. The goal is to find a solution that is feasible for almost all possible uncertainty realizations and maximizes the expectation of some function of the decisions variables.


The field of probabilistic methods \cite{TemIsh:07,Calafiore11,Tempo13} has received a growing attention in the systems and control community. Two complementary approaches, non-sequential and sequential,
have been proposed. A classical approach for non-sequential methods is based
upon statistical learning theory \cite{Vapnik98}, \cite{Vidyasagar97}. Subsequent work
along this direction includes \cite{Koltchinskii00}, \cite{Vidyasagar01a}, \cite{Vidyasagar01b}, \cite{Alamo09},
\cite{CDTVW:14}. Furthermore, in \cite{Alamo10ACC}, \cite{Alamo10} and \cite{Luedtke08} the case in which the design parameter set has finite cardinality is analyzed. The advantage of these methods is that the problem
under attention may be non-convex. For convex optimization problems, a  non-sequential
paradigm, denoted as the scenario approach, has been introduced
in \cite{Calafiore05} and \cite{Calafiore06}, see also \cite{Campi08}, \cite{Campi11}, \cite{CalafioreSIAM10}, \cite{Alamo10ACC} for more advanced results, and \cite{PrGaVi:14}, \cite{VaKuRu:12}
for recent developements in the areas of stochastic hybrid systems and multi-stage optimization, respectively. Finally, we refer to
\cite{FarVan:03}
for a randomized approach to solve approximate dynamic programming.

In non-sequential methods, the original robustness
problem is reformulated as a single optimization
problem with sampled constraints, which are randomly generated. A relevant feature of these methods is that they do not require any validation step and the sample complexity is defined a priori. The main result of this line of research is to derive explicit
 lower bounds to the required sample size. However, the obtained explicit sample bounds can be overly conservative because they rely on a worst-case analysis and grow (at least linearly) with the number of decision variables.

For sequential methods, the resulting iterative algorithms
are based on stochastic gradient \cite{Calafiore01}, \cite{Polyak01}, ellipsoid
iterations  \cite{Kanev03}, \cite{Oishi07};
or analytic center cutting plane methods
\cite{Calafiore07}, \cite{Takayuki1},  see also \cite{Alamo07,Chamanbaz13} for other classes of sequential algorithms.
Convergence properties in finite-time are one of the focal
points of these papers. Various control problems have been
solved using these sequential randomized algorithms, including
robust LQ regulators \cite{Polyak01}, switched systems \cite{Liberzon04} and uncertain linear
matrix inequalities (LMIs) \cite{Calafiore01}. Sequential methods are often
used for uncertain convex feasibility problems because  the computational effort at each iteration is affordable.
However, they have been studied also for non-convex problems, see \cite{Alamo09}, \cite{Ishii05}.

The common feature of most of these sequential algorithms is the use of the validation strategy presented in \cite{Oishi07} and \cite{Dabbene10}. The candidate solutions provided at each iteration of these algorithms are tested using a validation set which is drawn according to the probability measure associated to the uncertainty. If the candidate solution satisfies the design specifications for every sampled element of this validation set, then it is classified as probabilistic solution and the algorithm terminates. The main point in this validation scheme is that the cardinality of the validation set increases very mildly at each iteration of the algorithm. The strategy guarantees that, if a probabilistic solution is obtained, then it meets some probabilistic specifications.

In this paper, we derive the sample complexity for various analysis and design problems related to uncertain systems.
In particular we provide new results which guarantee that the tail of the binomial distribution is bounded by a pre-specified value.
These results are then applied to the analysis of worst-case performance and constraint violation. With regard to design problems, we consider the special cases of finite families and  robust convex optimization problems. This contribution improves the existing results by reducing the number of samples required to solve the design problem. We remark that the results we have obtained are fairly general and the assumptions on convexity and on finite families appear only in Section 4 which deals with probabilistic analysis and design.

The second main contribution of this paper is to propose a sequential validation scheme, denoted as Sequentially Probabilistic Validation (SPV), which allows the candidate solution to violate the design specifications for one (or more) of the members of the validation set. The idea of allowing some violations of the constraints is not new and can be found, for example, in the context of system identification \cite{Bai02}, chance-constrained optimization \cite{Campi11} and statistical learning theory \cite{Alamo09}. This scheme makes sense in the presence of soft constraints or when a solution satisfying the specifications for all the admissible uncertainty realizations can not be found. In this way, we improve the existing results with this relaxed validation scheme that reduces the chance of not detecting the solution even when it exists. Furthermore, we also show that a strict validation scheme may not be well-suited for some robust design problems.

This paper is based on the previous works of the authors \cite{Alamo10ACC} and \cite{Alamo12}. However, some results are completely new (Property 4) and others (Theorem 2, Property 1 and Property 3 and their proofs) are significant improvements of the preliminary results presented in the conference papers. Furthermore, the unifying
approach studied here, which combines sample complexity results with SPV algorithms, was not present in previous papers. Finally, the numerical example
in Section \ref{sec:applications}, which compares various approaches available in the literature, is also new.  The rest of the paper is organized as follows. In the next section, we first introduce the problem formulation. In Section \ref{sec:explicit:binomial}, we provide bounds for the binomial distribution which are used in
Section \ref{sec:analysis:design} to analyze the probabilistic properties of different schemes in\-vol\-ving randomization. In Section \ref{sec:SPValgorithms}, we introduce the proposed family of probabilistically validated algorithms. The sample complexity of the validating sets is analyzed in Section \ref{sec:samplesize}.  A detailed comparison with the validation scheme presented in \cite{Oishi07} is provided in Section \ref{sec:Oishi}.
A numerical example where different schemes are used to address a robust identification problem is presented in Section \ref{sec:applications}. The paper ends with a section of conclusions and an appendix which contains some auxiliary properties and proofs that are used in the previous sections.

\section{Problem Statement}\label{sec:ProblemStatement}

We assume that a probability measure ${\rm Pr}_{\setW}$ over the
sample space $\setW$ is given.  Given $\setW$, a collection of $N$
independent identically distributed (i.i.d.) samples
$\c{w}=\{w^{(1)},\ldots,w^{(N)}\}$ drawn from $\setW$ belongs to the Cartesian product $\setW^N=\setW\times \cdots \times
\setW$ ($N$ times). Moreover, if the collection $\c{w}$ of $N$
i.i.d. samples $\{w^{(1)},\ldots,w^{(N)}\}$ is generated from
$\setW$ according to the probability measure ${\rm Pr}_{\setW}$,
then the {\it multisample} $\c{w}$ is drawn according to the
probability measure ${\rm Pr}_{\setW^N}$. The scalars $\eta\in(0,1)$ and $\delta \in(0,1)$ denote probabilistic parameters called accuracy and confidence, respectively. Furthermore,
$\ln(\cdot)$ is the natural logarithm and ${\rm e}$ is the Euler number. For $x\in\mathds{R}$, $x\geq 0$, $\lfloor  x \rfloor$ denotes the largest integer smaller than or equal to $x$;  $\lceil x \rceil$ denotes the smallest integer greater or equal than $x$.
For  $\alpha>1$,
\[\xi(\alpha):=\Sum{k=1}{\infty}\frac{1}{k^\alpha}\] denotes the Riemann zeta function.

In a robustness problem,  the controller parameters and auxiliary variables are parameterized by means of
a decision variable vector $\theta$, which is denoted as
{\it design parameter}
and is restricted to a set $\Theta$. Furthermore, the uncertainty $w$ is bounded in the set $\setW$ and represents one of the admissible
uncertainty realizations. We also consider a binary measurable
function $g:\Theta\times \setW\to \{0,1\}$ and a  real measurable
function $f:\Theta\times \setW \to \mathds{R}$ which helps to formulate
the specific design problem under attention. More precisely,
the binary function $g:\Theta\times \setW\to \{0,1\}$, is defined
as
 \[g(\theta,w):=\bsis{rcl} 0 & &\mbox{if } \theta \mbox{ meets design specifications for  } w \\ 1 & & \mbox{otherwise}, \esis \]
where design specifications are, for example, $H_\infty$ norm bounds on the sensitivity function, see specific examples in \cite{Tempo13}, or the numerical example in Section \ref{sec:applications}.

Given $\theta\in \Theta$, the constraint $g(\theta,w)=0$ is satisfied for a subset of $\setW$. This concept is rigorously formalized by means of the
notion of {\it \nameE}, which is now introduced.

\begin{definition} \label{def:prob:violation} [{\nameE}]
Consider a probability measure ${\rm Pr}_{\setW}$ over $\setW$ and
let $\theta\in \Theta$ be given. The {{\nameE}} of $\theta$ for
the function $g:\Theta\times\setW\to\{0,1\}$ is defined as
\[\pV(\theta):={\rm Pr}_{\setW}\,\{\,g(\theta,w)=1\,\}.\]
\end{definition}

Using this notion we study the robust optimization problem
\begin{equation}
\label{equ:robustproblem} \min\limits_{\theta\in \Theta} J(\theta) \;\; \mbox{subject to}\;\; E(\theta)\leq \eta,
\end{equation}
where $J:\Theta\to (-\infty,\infty)$ is a measurable function which represents the controller performance and $\eta \in (0,1)$ is a probabilistic accuracy. Given accuracy $\eta\in(0,1)$ and confidence $\delta\in(0,1)$, the main point of the probabilistic approach is to design an algorithm such that any probabilistic solution $\hat{\theta}$ obtained by running the algorithm, satisfies $E(\hat{\theta})\leq \eta$ with probability no smaller than $1-\delta$.

Even in analysis problems when $\theta\in\Theta$ is given, it is often very hard to compute the
exact value of the {\nameE} $\pV(\theta)$ because this requires to solve a multiple integral with a usually non-convex domain of integration. However, we can approximate its
value using the concept of empirical mean. For given
$\theta\in\Theta$, and multisample $\c{w} = \{ w^{(1)}, \ldots, w^{(N)}\}$ drawn according to the
probability measure ${\rm Pr}_{\setW^N}$, the empirical mean of $g(\theta,w)$ with
respect to $\c{w}$ is defined as
\[\hat{E}(\theta, \c{w}):=\frac{1}{N}
\sum\limits_{i=1}^{N} g(\theta,w^{(i)}).\]
Clearly, the empirical mean $\hat{E}(\theta, \c{w})$ is
a random variable. Since $g(\cdot,\cdot)$ is a binary function,
$\hat{E}(\theta, \c{w})$ is always within the closed interval
$[0,1]$.

The power of randomized algorithms stems from the fact that they can approximately solve non-convex design problems (with
no-violation) of the type
\begin{equation}\label{general:optimizationproblem}
\min\limits_{\theta\in \Theta} J(\theta) \;\; \mbox{subject to}\;
g(\theta,w)=0, \; \mbox{for all}\; w\in \setW.
 \end{equation}

In this setting, we draw $N$ i.i.d. samples $\{ w^{(1)}, \ldots, w^{(N)}\}$ from $\setW$ according to probability ${\rm Pr}_\setW$ and solve the sampled optimization problem
\begin{equation}\label{sampled:optimizationproblem}
\min\limits_{\theta\in \Theta} J(\theta) \;\; \mbox{subject to}\;
g(\theta,w^{(\ell)})=0, \; \ell=1,\ldots, N.
 \end{equation}
 Since obtaining a global solution to this problem is still a difficult task in general, in this paper we analyze the probabilistic properties of {\it any suboptimal} solution. Furthermore, if at most $m$ violations of the $N$ constraints are allowed, the following sampled problem can be used to obtain a probabilistic relaxation to the original problem (\ref{general:optimizationproblem})
\begin{equation}\label{problem:m}
\min\limits_{\theta\in \Theta} J(\theta) \;\; \mbox{subject to}\;
\Sum{\ell=1}{N} g(\theta,w^{(\ell)})\leq m.
 \end{equation}
Randomized strategies to solve problems (\ref{sampled:optimizationproblem}) and (\ref{problem:m}) have been studied in \cite{Alamo09}, see also \cite{Tempo13}. In order to analyze the probabilistic properties of any feasible solution to problem (\ref{problem:m}), we introduce the definitions of
non-conforming feasible set and probability of failure.

\begin{definition}\label{def:nonconformingset}[{non-conforming feasible set}]
Given $N$, the integer $m$ where $0\leq m < N$, $\eta\in(0,1)$, $g:\Theta\times
\setW\to \{0,1\}$ and multisample $\c{w} = \{ w^{(1)}, \ldots, w^{(N)}\}$, drawn according to the
probability measure ${\rm Pr}_{\setW^N}$,  the non-conforming feasible set $\Theta(\c{w},\eta,m)$ is defined as
\[\Theta(\c{w},\eta,m):=\set{\theta\in \Theta}{\hat{E}(\theta,\c{w})\leq \frac{m}{N}\, \,\mbox{ \rm and}\,\,\pV(\theta) >\eta}.\]
\end{definition}

\begin{definition}\label{prob:failure:feasible}[{\namep}]
Given $N$, the integer $m$ where $0\leq m < N$, $\eta\in(0,1)$ and $g:\Theta\times
\setW\to \{0,1\}$, the {\namep}, denoted by $p(N,\eta,m)$
is defined as
\[ p(N,\eta,m):=  {\rm Pr}_{\setW^N} \{\, \Theta(\c{w},\eta,m) \;\mbox{\rm is not empty} \}.\]
\end{definition}
The probability $p(N,\eta,m)$ defined here is slightly different
than the probability of one-sided constrained failure introduced
in \cite{Alamo09}.  We notice that the non-conforming feasibility set is empty with probability $1-p(N,\eta,m)$. This means that every feasible solution $\theta\in\Theta$ to problem (\ref{problem:m}) satisfies $E(\theta)\leq \eta$ with probability $1-p(N,\eta,m)$. Given the confidence parameter $\delta\in(0,1)$, the objective is to obtain explicit expressions yielding a minimum number of samples $N$ such that $p(N,\eta,m)\leq \delta$.

\section{Sample complexity for the binomial distribution}\label{sec:explicit:binomial}

In this section, we provide bounds for the binomial distribution which are used in
Section \ref{sec:analysis:design}. Given a positive integer $N$ and a nonnegative integer $m$, $m\leq
N$, and $\eta\in (0, 1)$, the binomial distribution function is given by
\begin{eqnarray}
B(N,\eta,m) := \Sum{i=0}{m}\conv{N}{i}\eta^i (1-\eta)^{N-i}. \nonumber
\end{eqnarray}
The problem we address in this section is the explicit computation of the {\it sample complexity}, i.e. a function $\tilde{N}(\eta,m,\delta)$ such that the inequality $B(N,\eta,m)\leq \delta$ holds for any $N\geq \tilde{N}(\eta,m,\delta)$, where $\delta\in(0,1)$.  As it will be illustrated in the following section, the inequality $B(N,\eta,m)\leq \delta$ plays a fundamental role in probabilistic methods. Although some explicit expressions are available, e.g. the multiplicative and additive forms of Chernoff bound \cite{Chernoff52}, the results
 obtained in this paper are tuned on the specific inequalities stemming from the problems described in Section \ref{sec:analysis:design}.

The following technical lemma provides an upper bound for the binomial distribution
$B(N,\eta,m)$.
\begin{lemma}\label{property:a}
Suppose that $\eta\in(0,1)$ and that the nonnegative integer $m$
and the positive integer $N$ satisfy $m\leq N$. Then,
$
 B(N,\eta,m) \leq  a^m\left(\frac{\eta}{a}+1-\eta\right)^{N},\;\; \forall a\geq 1.
$
\end{lemma}

\ourproof
The proof of the lemma follows from the following sequence of inequalities:
\begin{eqnarray*}
B(N,\eta,m) &= & a^m \Sum{i=0}{m}\conv{N}{i} a^{-m} \eta^i(1-\eta)^{N-i} \\
  &\leq &  a^m \Sum{i=0}{m} \conv{N}{i} a^{-i} \eta^i(1-\eta)^{N-i} \\  & \leq & a^m \Sum{i=0}{N} \conv{N}{i} \left(\frac{\eta}{a}\right)^i(1-\eta)^{N-i} \\
&=& a^m\left(\frac{\eta}{a}+1-\eta\right)^{N}.
\end{eqnarray*}
\hfill $~$  \qed

We notice that each particular choice of $a\geq 1$ provides an
upper bound for $B(N,\eta,m)$. When using Lemma \ref{property:a}
to obtain a specific sample complexity, the selected value for
$a$ plays a  significant role.

\begin{lemma}\label{property:a:geq:one}
Given $\delta\in(0,1)$ and the nonnegative integer $m$, suppose that the integer $N$ and the scalars  $\eta\in(0,1)$ and $a>1$ satisfy the inequality
\begin{equation}\label{ineq:a}
 N\geq \frac{1}{\eta}\left(\frac{a}{a-1}\right)\left(\ln\,\frac{1}{\delta}+m\ln\, a\right). \label{eq5} \end{equation}
Then, $m < N$ and $ B(N,\eta,m)  \leq \delta$.
\end{lemma}

\ourproof
We first prove that if inequality (\ref{ineq:a}) is satisfied then $m<N$.
 Since $\eta\in (0,1)$ and $\delta \in(0,1)$,  (\ref{ineq:a}) implies
\[N > \left(\frac{a}{a-1}\ln\,a\right)m.\] Next, we notice that
\[ \frac{\rm{d}}{{\rm{d}}a}\left( \;\frac{a}{a-1}\ln\,a \right)= \left(\frac{-1}{(a-1)^2}\right)\ln\,a
+\frac{1}{a-1}.\]
Since $\ln\,a < a-1$ for every $a>1$, it follows that
\[ \frac{\rm{d}}{{\rm{d}}a}\left( \;\frac{a}{a-1}\ln\,a \right) > \left(\frac{-1}{(a-1)^2}\right)(a-1)
+\frac{1}{a-1}=0.\]
Using this fact, we conclude that $\frac{a}{(a-1)}\ln\,a$
is a strictly increasing function for $a>1$. This means that
\[N > \left(\frac{a}{a-1}\ln\,a\right)m \geq \lim\limits_{\hat{a}\to 1}
\left(\frac{\hat{a}}{\hat{a}-1}\ln\,\hat{a}\right)m =m.\] We now
prove that (\ref{ineq:a}) guarantees that
$a^m(\frac{\eta}{a}+1-\eta)^N \leq \delta$. The inequality
(\ref{ineq:a}) can be rewritten as
\begin{equation}\label{ineq:b}
N\eta \left(\frac{a-1}{a}\right) \geq \ln\,\frac{1}{\delta}+m\ln\, a. \end{equation}
Since $x\leq -\ln\,(1-x)$ for every $x\in(0,1)$, and $\eta(\frac{a-1}{a})\in(0,1)$, from inequality (\ref{ineq:b}), we obtain a sequence of inequalities
\[-N\ln\,\left(1-\eta \left(\frac{a-1}{a}\right)\right) \geq \ln\,\frac{1}{\delta}+m\ln\, a\]
\[ \ln\,\delta \geq  m\ln\, a+ N\ln\,\left(1-\eta\left(\frac{a-1}{a}\right)\right) \]
\[\delta \geq a^m\left(\frac{\eta}{a}+1-\eta\right)^N .\]
We have therefore proved that inequality (\ref{ineq:a}) implies
$m\leq N$ and $a^m(\frac{\eta}{a}+1-\eta)^N \leq \delta$. The
claim of the property follows directly from Lemma
\ref{property:a}. \hfill $~$  \qed

Obviously, the best sample size bound is obtained taking the infimum with respect to $a>1$. However, this requires to solve numerically a one-dimensional optimization problem for given $\eta$, $\delta$ and $m$. We observe that a suboptimal value  can be immediately obtained setting $a$ equal to the Euler constant, which yields the sample complexity
\begin{equation}\label{equ:linearEuler}
 N \geq \frac{1}{\eta}\left(\frac{\ee}{\ee-1}\right)\left(\ln\,\frac{1}{\delta}+m\right).
 \end{equation}
 Since $\frac{\ee}{\ee-1}< 1.59$, we obtain $ N \geq \frac{1.59}{\eta} \left(\ln\,\frac{1}{\delta}+m\right)$,
which is (numerically) a significant improvement of the bound given in
\cite{CalafioreSIAM10} and other bounds available in the literature
\cite{Calafiore11}. We also notice that, if $m>0$ then the choice
\[a=1+\frac{\ln\,\frac{1}{\delta}}{m}+ \sqrt{2\frac{\ln\, \frac{1}{\delta} }{m}}\]
 provides a less conservative bound at the price of a more involved expression \cite{Alamo10ACC}. Based on extensive numerical computations for several values of $\eta$, $\delta$ and $m$ we conclude that this bound is very close to the ``optimal'' one. Note, however, that the optimal value can be obtained numerically using the Lambert $W$ function \cite{Corless96}. In the next corollary, we present another more involved sample complexity bound which improves (\ref{equ:linearEuler}) for some values of the parameters.

\begin{corollary}\label{corollary:sqrt}
Given $\delta\in(0,1)$ and the nonnegative integer $m$, suppose that the integer $N$ and the scalar  $\eta\in(0,1)$ satisfy the inequality
\begin{equation}\label{ineq:sqrt}
 N \geq \frac{1}{\eta}\left(m+\ln\,\frac{1}{\delta}+\sqrt{2m\ln\,\frac{1}{\delta}}\, \right).\end{equation}
Then, $m<N$ and $B(N,\eta,m) \leq \delta$.
\end{corollary}
 The proof of this corollary is shown in the appendix.

\section{Sample complexity for probabilistic analysis and design}\label{sec:analysis:design}

We now study some problems in the context of randomized algorithms where one encounters inequalities of the form
$B(N,\eta,m) \leq
\delta. $
In particular, we show how the results of the previous section can be used to derive explicit sample size bounds which guarantee that the probabilistic solutions obtained from different randomized approaches meet some pre-specified probabilistic properties.

In Subsection \ref{subsec:finite} we derive bounds on $p(N,\eta,m)$ when $\Theta$ consists of a finite number of elements. On the other hand, if $\Theta$ consists of an infinite number of elements, a deeper analysis involving statistical learning theory is needed   \cite{Tempo13}, \cite{Vidyasagar97}.  In Subsection \ref{subsec:scenario} we study the probabilistic properties of the optimal solution of problem (\ref{sampled:optimizationproblem}) under the assumption that $g(\theta,w)=0$ is equivalent to $f(\theta,w)\leq 0$, where $f:\Theta\times \setW\to \mathds{R}$ is a convex function with respect to $\theta$ in $\Theta$. In this case, the result is not expressed in terms of probability of failure because it applies only to the optimal solution of problem (\ref{sampled:optimizationproblem}), and not to every feasible solution.

\subsection{Worst-case performance analysis}

We recall a result shown in \cite{TeBaDa:97} for the probabilistic worst-case performance analysis.
\begin{theorem}
Given the function $f:\Theta\times \setW\to \mathds{R}$ and $\hat{\theta}\in\Theta$, consider the multisample
$\c{w}=\{ w^{(1)}, \ldots, w^{(N)}\}$ drawn from $\setW^N$
according to probability ${\rm Pr}_{\setW^N}$ and define
$ \gamma=\max\limits_{\ell=1,\ldots,N} f(\hat{\theta},w^{(\ell)}).$
If
\[ N\geq \frac{\ln\,\frac{1}{\delta}}{\ln\,\frac{1}{1-\eta}},\]
then ${\rm Pr}_{\setW} \{ w\in \setW \,:\, f(\hat{\theta},w)>\gamma \}\leq \eta$ with probability no smaller than $1-\delta$.
\end{theorem}

The proof of this statement can be found in \cite{TeBaDa:97} and is based on the fact that  ${\rm Pr}_{\setW} \{ w\in \setW \,:\, f(\hat{\theta},w)>\gamma \}\leq \eta$ with probability no smaller than $1-(1-\eta)^N$. Therefore, it suffices to take $N$ such that
$B(N,\eta,0)=(1-\eta)^N\leq \delta$.

\subsection{Finite families for design}\label{subsec:finite}

 We consider the non-convex sampled problem (\ref{problem:m}) for the special case when $\Theta$ consists of a set of finite cardinality $n_C$.   As a motivation, we study the case when, after an appropriate normalization procedure,
 the design parameter set is rewritten as
$\hat{\Theta}=\set{\theta\in\mathds{R}^{\nt}}{\| \theta\|_{\infty}\leq
1}.$ Suppose also that a gridding approach is adopted. That is, for each
component $\theta_j$, $j=1,\ldots,\nt$ of the design parameters
$\theta\in\mathds{R}^{\nt}$, only $n_{C_j}$ equally spaced values are
considered. That is, $\theta_j$ is constrained into the set
$\Upsilon_j=\set{-1+\frac{2(t-1)}{(n_{C_j}-1)}}{t=1,\ldots,n_{C_j}}.$
With this gridding procedure, the following finite cardinality set
$\Theta=\set{[\theta_1,\ldots,\theta_\nt]^T}{\theta_j\in
\Upsilon_j, \; j=1,\ldots, \nt}$ is obtained. We notice that the
cardinality of the set is $n_C=\prod_{j=1}^{\nt} n_{C_j}$. Another
situation in which the finite cardinality assumption holds is when
a finite number of random samples in the space of design parameter
are drawn according to a given probability, see e.g.
\cite{Fujisaki06,Vidyasagar01a}.

The following theorem states the relation between the binomial
distribution and the probability of failure under this finite
cardinality assumption.

\begin{theorem}\label{theorem:finite}
Suppose that the cardinality of $\Theta$ is no larger than $n_C$, $n_C>0$, $\eta\in(0,1)$ and $m<N$.
Then,
\[p(N,\eta,m)<  n_{C}B(N,\eta,m).\]
\end{theorem}

\ourproof
If there is no element in $\Theta$ with probability of violation larger than $\eta$, then the non-conforming feasible set is empty for every multisample $\c{w}$ and $p(N,\eta,m)=0<n_C B(N,\eta,m)$.

Suppose now that the subset of $\Theta$ of elements with probability of violation larger than $\eta$ is not empty. Denote $\{ \theta^{(1)},\theta^{(2)},\ldots, \theta^{(\tilde{n})}\}$ such a set. In this case, given a multisample $\c{w}$, the non-conforming feasible set is not empty if and only if the empirical mean is smaller or equal than $\frac{m}{N}$ for at least one of the elements of this set. Therefore
\begin{eqnarray*}
p(N,\eta,m) &=& {\rm Pr}_{\setW^N} \{\, \Theta(\c{w},\eta,m) \;\mbox{is not empty} \}\\
&=&  {\rm Pr}_{\setW^N} \{\, \min\limits_{1\leq k \leq \tilde{n}} \hat{E}(\theta^{(k)},\c{w}) \leq \frac{m}{N}  \}\\
&\leq & \Sum{k=1}{\tilde{n}}  {\rm Pr}_{\setW^N} \{\, \hat{E}(\theta^{(k)},\c{w}) \leq \frac{m}{N}  \}\\
& = & \Sum{k=1}{\tilde{n}}  B(N,E(\theta^{(k)}),m)\\
& < & \Sum{k=1}{\tilde{n}}  B(N,\eta,m) = \tilde{n} B(N,\eta,m).
\end{eqnarray*}
Notice that the last inequality is due to the fact that $E(\theta^{(k)}) >\eta$, $k=1,\ldots,\tilde{n}$ and that the binomial distribution is a strictly decreasing function of $\eta$ if $m<N$ (see Property \ref{prop:decreasingBinomial} in the Appendix). To conclude the proof it suffices to notice that $\tilde{n}\leq n_C$.
\qed

Consider now the optimization problem
(\ref{problem:m}). It follows from Lemma
\ref{theorem:finite} that to guarantee that every
feasible solution $\hat{\theta}\in \Theta$ satisfies
$E(\hat{\theta})\leq \eta$ with probability no smaller than
$1-\delta$, it suffices to take $N>m$ such that $n_C B(N,\eta,m)\leq
\delta$, where $n_C$ is an upper bound on the cardinality of
$\Theta$. As it will be shown next, the required sample complexity
in this case grows with the logarithm of $n_C$. This means
that we can  consider finite families with high cardinality and
still obtain very reasonable sample complexity bounds.

\begin{theorem}\label{theorem:finite:explicit}
Suppose that the cardinality of $\Theta$ is no larger than
$n_{C}$. Given the nonnegative integer $m$,
$\eta\in(0,1)$ and $\delta\in(0,1)$,  if

\begin{equation}\label{teoremafinito}
N \geq \inf\limits_{a>1}\,\frac{1}{\eta}\left(\frac{a}{a-1}\right)\left(\ln\,\frac{n_{C}}{\delta}+m\ln\, a\right) \end{equation}
then $p(N,\eta,m)\leq \delta$. Moreover, if
\[N \geq \frac{1}{\eta}\left(m+\ln\,\frac{n_{C}}{\delta}+\sqrt{2m\ln\frac{n_{C}}{\delta}}\, \right)\]
 then $p(N,\eta,m)\leq \delta$.
\end{theorem}

 \ourproof
From Lemma \ref{theorem:finite} we have that $p(N,\eta,m)\leq \delta$ provided that $m<N$ and $B(N,\eta,m)$ $\leq \frac{\delta}{n_C}$. The two claims of the property now follow directly from Lemma \ref{property:a:geq:one} and Corollary \ref{corollary:sqrt} respectively.

\hfill $~$  \qed

From the definition of $p(N,\eta,m)$ and Theorem \ref{theorem:finite:explicit} we conclude that if one draws $N$ i.i.d. samples $\{w^{(1)},\ldots,w^{(N)}\}$ from $\setW$ according to probability $\rm{Pr}_{\setW}$, then with probability no smaller than $1-\delta$, all the feasible solutions to problem (\ref{problem:m}) have a probability of violation no larger than $\eta$, provided that the cardinality of $\Theta$ is upper bounded by $n_C$ and the sample complexity is given by
\[ N \geq \frac{1}{\eta}\left(m+\ln\,\frac{n_{C}}{\delta}+\sqrt{2m\ln\frac{n_{C}}{\delta}}\, \right).\]
We remark that taking $a$ equal to the Euler constant
in (\ref{teoremafinito}), the following sample size bound
\[ N \geq \frac{1}{\eta}\left(\frac{\ee}{\ee-1}\right)\left(\ln\,\frac{n_{C}}{\delta}+m\right)\]
is immediately obtained from Theorem \ref{theorem:finite:explicit}.
If $m>0$ then a suboptimal value for $a$ is given by \[a=1+\frac{\ln\,\frac{n_{C}}{\delta}}{m}+ \sqrt{2\frac{\ln\, \frac{n_{C}}{\delta} }{m}}.\]

\subsection{ Optimal robust optimization for design}\label{subsec:scenario} In this subsection, we study
the so-called scenario approach for robust control introduced in
\cite{Calafiore06}. To address the semi-infinite optimization problem (\ref{general:optimizationproblem}), we solve the randomized  optimization problem (\ref{sampled:optimizationproblem}).  That is, we generate $N$ i.i.d. samples $\{ w^{(1)}, \ldots, w^{(N)}\}$ from $\setW$ according to the probability ${\rm Pr}_\setW$ and then solve the following sampled optimization problem:
\begin{equation}\label{convex:optimizationproblem}
\min\limits_{\theta\in \Theta} J(\theta) \;\; \mbox{subject to}\;
g(\theta,w^{(\ell)})=0, \; \ell=1,\ldots, N.
 \end{equation}
We consider here the particular case in which $J(\theta)=c^T \theta$,
 the constraint $g(\theta,w)=0$ is convex in $\theta$ for all $w\in \setW$ and the solution of (\ref{convex:optimizationproblem}) is
 unique. 
These assumptions are now stated precisely.
\begin{assumption}\label{assumption:convex} {\rm [convexity] } Let $\Theta\subset \mathds{R}^{\nt}$ be
a convex and closed set. We
assume that
\[J(\theta):=c^T \theta \;\; \mbox{ and } \;\;
 g(\theta,w):=\bsis{rl} 0 & \mbox{if}\;f(\theta,w)\leq 0,
 \\
1&\mbox{otherwise}\esis \] where $f:\Theta\times\setW
\to [-\infty,\infty]$ is convex in $\theta$ for every fixed value of
$w\in \setW$.
\end{assumption}

\begin{assumption}\label{assumption:unique}{\rm [feasibility and uniqueness]}
For all possible multisample extractions $\{w^{(1)}$, $\ldots$, $w^{(N)}\}$, the optimization problem (\ref{convex:optimizationproblem}) is always feasible and attains a unique optimal solution. Moreover, its feasibility domain has a nonempty
interior. \end{assumption}

Uniqueness may be assumed essentially without loss of generality, since in case of multiple optimal solutions one may always introduce a suitable
tie-breaking rule \cite{Calafiore06}.
We now state a result that relates the
binomial distribution to the probabilistic properties of the optimal solution obtained from (\ref{convex:optimizationproblem}).
See \cite{Campi08,CalafioreSIAM10,Campi11}.

\begin{lemma}\label{Campi:result} Let Assumptions 1 and 2 hold. Suppose that $N$, $\eta\in(0,1)$ and $\delta\in(0,1)$ satisfy the inequality
\begin{equation}\label{inequality}
\Sum{i=0}{\nt-1}\conv{N}{i}\eta^i(1-\eta)^{N-i}\leq \delta.
\end{equation}  Then, with probability no smaller than $1-\delta$,
the optimal solution $\hat{\theta}_N$ to the optimization problem
(\ref{convex:optimizationproblem}) satisfies the inequality
$E(\hat{\theta}_N)\leq \eta$.
\end{lemma}

We now state an explicit sample size bound, which improves upon previous bounds, to guarantee that the probability of violation is smaller than $\eta$ with probability at least $1-\delta$.

\begin{theorem}\label{theorem:convexscenario}
 Let Assumptions 1 and 2 hold. Given $\eta\in(0,1)$ and $\delta\in(0,1)$, if
\begin{equation}\label{ineq:scenario:uno}
 N \geq \inf\limits_{a>1} \frac{1}{\eta}\left(\frac{a}{a-1}\right)\left(\ln\,\frac{1}{\delta}+(\nt-1)\ln\, a\right)
 \end{equation}
or
\begin{equation}\label{ineq:scenario:dos}
 N \geq \frac{1}{\eta}\left(\ln\,\frac{1}{\delta}+(\nt-1)+\sqrt{2(\nt-1)\ln\frac{1}{\delta}}\, \right)
\end{equation}
then, with probability no smaller than $1-\delta$,
the optimal solution $\hat{\theta}_N$ to the optimization problem
(\ref{convex:optimizationproblem}) satisfies the inequality
$E(\hat{\theta}_N)\leq \eta$.
\end{theorem}

\ourproof
From Lemma \ref{Campi:result} it follows that it suffices to take $N$ such that $B(N,\eta,\nt-1)\leq \delta$. Both inequalities (\ref{ineq:scenario:uno}) and (\ref{ineq:scenario:dos}) guarantee that $B(N,\eta,\nt-1)\leq \delta$ (see  Lemma \ref{property:a:geq:one}  and Corollary \ref{corollary:sqrt} respectively). This completes the proof.

\hfill $~$  \qed

Taking $a$ equal to the Euler constant in (\ref{ineq:scenario:uno}), we obtain
\[ N \geq \frac{1}{\eta}\left(\frac{\ee}{\ee-1}\right)\left(\ln\,\frac{1}{\delta}+\nt-1\right)\]
which improves the bound given in \cite{CalafioreSIAM10} and other bounds available in the literature \cite{Calafiore11}. More precisely, the constant 2 appearing in \cite{CalafioreSIAM10} is reduced to $\frac{\mathrm{e}}{(\mathrm{e} -1)} \approx 1.59$, which is (numerically) a substantial improvement for small values of $\eta$.
If $\nt>1$ a suboptimal value for $a$ is given by
\emph{}\[a=1+\frac{\ln\,\frac{1}{\delta}}{\nt-1}+
\sqrt{2\frac{\ln\, \frac{1}{\delta} }{\nt-1}}.\]


\section{Sequential algorithms with probabilistic validation}\label{sec:SPValgorithms}

In this section, we present a general family of randomized algorithms, which we denote as {\it Sequential Probabilistic Validation} (SPV) algorithms. The main feature of this class of algorithms is that they are based on a probabilistic validation step. This family includes most of the sequential randomized algorithms that have been presented in the literature and are discussed in the introduction of this paper.  

Each iteration of an SPV algorithm includes the computation of a candidate solution for the problem and a subsequent validation step. The results provided in this paper are basically independent of the particular strategy chosen to obtain candidate solutions. Therefore, in the following discussion we restrict ourselves to a generic candidate solution. The accuracy $\eta\in(0,1)$ and confidence $\delta\in(0,1)$ required for the probabilistic solution play a relevant role when determining the sample size of each validation step. The main purpose of this part of the paper is to provide a validation scheme which guarantees that, for given accuracy $\eta$ and confidence $\delta$, all the probabilistic solutions obtained running the SPV algorithm have a probability of violation no larger than $\eta$ with probability no smaller than $1-\delta$.


We enumerate each iteration of the algorithm by means of an integer $k$. We denote by $m_k$ the number of violations that are allowed at the validation step of iteration $k$. We assume that $m_k$ is a function of $k$, that is,  $m_k=m(k)$ where the function $m:\mathds{N}\to \mathds{N}$ is given. We also denote by $M_k$ the sample size of the validation step of iteration $k$. We assume that $M_k$ is a function of $k$, $\eta$ and $\delta$. That is, $M_k=M(k,\eta,\delta)$ where $ M:\mathds{N}\times \mathds{R} \times \mathds{R} \to \mathds{N}$ has to be appropriately designed in order to guarantee the probabilistic properties of the algorithm. In fact, one of the main contributions of \cite{Oishi07,Dabbene10} is to provide this function for the particular case $m_k=0$ for every $k\geq 1$. The functions $m(\cdot)$ and $M(\cdot,\cdot,\cdot)$ are denoted as {\it level function} and {\it cardinality function} respectively.

We now introduce the structure of an SPV algorithm

\blista
\item Set accuracy $\eta\in(0,1)$ and confidence $\delta\in(0,1)$ equal to the desired levels. Set $k$ equal to 1.
\item Obtain a candidate solution $\hat{\theta}_k$ to the robust optimization problem (\ref{equ:robustproblem}).
\item  Set $m_k=m(k)$ and  $M_k= M(k,\eta,\delta)$.
 \item Obtain validation set $\setV_k =\{v^{(1)},\ldots,v^{(M_k)}\}$ drawing $M_k$ i.i.d. validation samples from $\setW$ according to probability ${\rm Pr}_{\setW}$.
\item If  $\Sum{\ell=1}{M_k} g(\hat{\theta}_k,v^{(\ell)})\leq m_k$, then $\hat{\theta}_k$ is a probabilistic solution.
\item Exit if the exit condition is satisfied.
\item $k=k+1$. Goto (ii).
\elista

Although the exit condition can be quite general, a reasonable choice is to exit after a given number of candidate solutions have been classified as probabilistic solutions or when a given computational time has elapsed since the starting of the algorithm. After exiting one could choose the probabilistic solution which maximizes a given performance index. We notice that in step (iv) we need to satisfy the i.i.d. assumption, and therefore sample reuse techniques are not applicable. In the next section, we propose a strategy to choose the cardinality of the validation set at iteration $k$ in such a way that, with probability no smaller than $1-\delta$, all candidate solutions classified as probabilistic solutions by the algorithm meet the accuracy $\eta$.

\section{Adjusting the validation sample size } \label{sec:samplesize}

The cardinality adjusting strategy provided in this section constitutes a generalization of that presented in \cite{Oishi07} and \cite{Dabbene10}. To obtain the results of this section we rely on some contributions on the sample complexity presented in the previous sections.

We now formally introduce the {\it failure function}.

\begin{definition}[failure function]
The function $\mu:\mathds{N}\to\mathds{R}$ is said to be a failure function if it satisfies the following conditions:
\blista \item $\mu(k)\in (0,1)$ for every positive integer $k$.
\item $\Sum{k=1}{\infty} \mu(k) \leq 1.$
\elista
\end{definition}

We notice that the function
\[\mu(k)=\frac{1}{\xi(\alpha)k^\alpha},\]
where $\xi(\cdot)$ is the Riemann zeta function,
is a failure function for every $\alpha>1$. This is due to the fact that $\Sum{i=1}{\infty} \frac{1}{k^\alpha}$ converges for every scalar $\alpha$ greater than 1 to $\xi(\alpha)$.  This family has been used in the context of validation schemes in \cite{Dabbene10} and in \cite{Oishi07} for the particular value $\alpha=2$.

\begin{property}\label{prop:bound}
Consider an SPV algorithm with given accuracy parameter $\eta\in(0,1)$, confidence $\delta\in(0,1)$, level function $m(\cdot)$ and cardinality function $M(\cdot,\cdot,\cdot)$. If $m(k)<M(k,\eta,\delta)$, for all $k\geq 1$, and there exists a failure function $\mu(\cdot)$ such that
\[ \Sum{i=0}{m(k)}\conv{M(k,\eta,\delta)}{i}\eta^i (1-\eta)^{M(k,\eta,\delta)-i} \leq \delta \mu(k), \;\; \forall k\geq 1\]
then, with probability greater than $1-\delta$, all the probabilistic solutions obtained running the SPV algorithm have a probability of violation no greater than $\eta$.
\end{property}

The proof of this property follows the same lines as the proof of Theorem 9 in \cite{Oishi07}.

\ourproof
We denote by $\delta_k$ the probability of  classifying at iteration $k$ the candidate solution $\hat{\theta}_k$ as a probabilistic solution under the assumption that the probability of violation $E(\hat{\theta}_k)$ is larger than $\eta$. Furthermore, let $M_k = M(k,\eta,\delta)$, then
\begin{eqnarray*}
\delta_k &=&  {\rm Pr}_{\setW^{M_k}} \{ \, \hat{E}(\hat{\theta}_k,\c{w}) \leq \frac{m_k}{M_k} \,\} \\
& = & \Sum{i=0}{m_k} \conv{M_k}{i} E(\hat{\theta}_k)^i (1-\hat{\theta}_k)^{M_k-i}  \\
& < & \Sum{i=0}{m_k} \conv{M_k}{i} \eta^i (1-\eta)^{M_k-i}.
\end{eqnarray*}
Property \ref{prop:decreasingBinomial} in the Appendix, $m_k< M_k$, and $E(\hat{\theta}_k)>\eta$ have been used to derive the last
inequality. Then, we obtain
\[\delta_k  < \Sum{i=0}{m(k)}\conv{M(k,\eta,\delta)}{i}\eta^i (1-\eta)^{M(k,\eta,\delta)-i} \leq \delta \mu(k).\]
  Therefore, the probability of misclassification of a candidate solution at iteration $k$ is smaller than $\delta\mu(k)$. We conclude that the probability of erroneously classifying one or more candidate solutions as probabilistic solutions is bounded by
\[\Sum{k=1}{\infty} \delta_k < \Sum{k=1}{\infty} \delta \mu(k) = \delta \Sum{k=1}{\infty} \mu(k) \leq \delta.\]
\hfill $~$  \qed

To design a cardinality function $M(\cdot,\cdot,\cdot)$ satisfying the conditions of Property \ref{prop:bound} we may use Corollary \ref{corollary:sqrt}.

We now present the main contribution of this part of the paper, which is a general expression for the cardinality of the validation set at each iteration of the algorithm.

\begin{theorem}\label{theo:maincontribution}
Consider an SPV algorithm with given accuracy $\eta\in(0,1)$, confidence $\delta\in(0,1)$ and level function $m(\cdot)$. Suppose also that $\mu(\cdot)$ is a failure function. Then, the cardinality function
\[ M(k,\eta,\delta) = \]
 \[\left\lceil \frac{1}{\eta} \left(m(k)+\ln\,\frac{1}{\delta \mu(k)}+\sqrt{2m(k)\ln\,\frac{1}{\delta \mu(k)}}\, \right) \right\rceil \]
guarantees that, with probability greater than $1-\delta$, all the probabilistic solutions obtained running the SPV algorithm have a probability of violation no greater than $\eta$.
\end{theorem}

\ourproof
Corollary \ref{corollary:sqrt} guarantees that the proposed choice for the cardinality function satisfies $m(k)<M(k,\eta,\delta)$, for all $k \geq 1$, and
\[ \Sum{i=0}{m(k)}\conv{M(k,\eta,\delta)}{i}\eta^i (1-\eta)^{M(k,\eta,\delta)-i} \leq \delta \mu(k), \; \forall k\geq 1.\]
The result then follows from a direct application of Property \ref{prop:bound}.
\hfill $~$  \qed

We notice that the proposed cardinality function $M(k,\eta,\delta)$ in Theorem \ref{theo:maincontribution} depends on the previous selection of the level function $m(\cdot)$ and the failure function $\mu(\cdot)$. A reasonable choice for these functions is $m(k)=\lfloor ak \rfloor$, where $a$ is a non-negative scalar and $\mu(k)= \frac{1}{\xi(\alpha)k^{\alpha}}$ where $\alpha$ is greater than one. We recall that this choice guarantees that $\mu(k)$ is a failure function. As shown in the following section, the proposed level and failure functions allow us to recover, for the particular choice $a=0$ the validation strategies proposed in \cite{Dabbene10} and \cite{Oishi07}. In the next corollary, we specify the generic structure of the SPV algorithm with the level function $m(k)=\lfloor ak \rfloor$, and state a probabilistic result.

\begin{corollary}\label{coro:algorithm}
Consider an SPV algorithm of the form given in Section \ref{sec:SPValgorithms} in which steps {\rm (i)} and
{\rm (iii)} are substituted by
{\rm
\blista
\item Set accuracy $\eta\in(0,1)$, confidence $\delta\in(0,1)$ and scalars $a\geq 0$, $\alpha>1$ equal to the desired levels. Set $k$ equal to 1.
\setcounter{enumi}{2}
\item  Set $m_k=\left\lfloor ak \right\rfloor $ and  \[M_k= \left\lceil \frac{1}{\eta} \left(m_k+\ln \frac{\xi(\alpha)k^{\alpha}}{\delta}+\sqrt{2m_k\ln\frac{\xi(\alpha)k^{\alpha}}{\delta}}\, \right) \right\rceil.\]
\elista
}
Then, with probability greater than $1-\delta$, all the probabilistic solutions obtained running the SPV algorithm have a probability of violation no greater than $\eta$.
\end{corollary}

\ourproof
The result is obtained directly from Theorem \ref{theo:maincontribution} using as level function $m(k)=\left\lfloor ak \right\rfloor $ and failure function $\mu(k)=\frac{1}{\xi(\alpha) k^{\alpha}}$.
\hfill $~$  \qed

Since the probabilistic properties of the algorithm presented in Corollary \ref{coro:algorithm} are independent of the particular value of $\alpha>1$, a reasonable choice for $\alpha$ is to select this parameter to minimize the cardinality of the validation sample set.

\section{Comparison with other validation schemes}\label{sec:Oishi}
In this section, we provide comparisons with the validation schemes presented in \cite{Oishi07,Dabbene10}. We notice that setting $a=0$ and $\alpha=2$ in Corollary \ref{coro:algorithm} we obtain $m(k)=0$ for every iteration $k$ and
\[M(k) =  \left\lceil \frac{1}{\eta}  \ln\,\left(\frac{\xi(2) k^2}{\delta} \right) \right\rceil =  \left\lceil \frac{1}{\eta}  \ln\,\left(\frac{\pi^2 k^2}{6\delta} \right) \right\rceil.\]
This is the same cardinality function presented in \cite{Oishi07} if one takes into account that for small values of $\eta$, $-\ln \, (1-\eta)$ can be approximated by $\eta$. In the same way, $a=0$ and $\alpha=1.1$ lead to the cardinality function presented in \cite{Dabbene10}.

We notice that not allowing any failure in each validation test makes perfect sense for convex problems if the feasibility set
$ \Theta_r = \set{\theta\in \Theta}{g(\theta,w)=0 \mbox{ for all } w\in \setW} $
is not empty. Under this assumption, the algorithm takes advantage of the validation samples that have not satisfied the specifications to obtain a new candidate solution. If $\Theta_r$ is not empty, a common feature of the methods which use this strict validation scheme is that a probabilistic solution  (not necessarily belonging to the feasibility set $\Theta_r$) is obtained in a finite number of iterations of the algorithm,  see e.g., \cite{Alamo07}, \cite{Calafiore07}, \cite{Oishi07}.

A very different situation is encountered when $\Theta_r$ is empty. We now state a property showing that a strict validation scheme ($a=0$) should not be used to address the case of empty robust feasible set because the algorithm might fail to obtain a probabilistic solution even if the set $\set{\theta\in\Theta}{E(\theta)\leq \eta}$ is not empty.

\begin{property}\label{property:never}
Consider the SPV algorithm presented in Corollary \ref{coro:algorithm} with $a=0$ and $\alpha>1$. Suppose that $E(\theta) \geq \mu >0$ for all $ \theta\in \Theta$. Then the SPV algorithm does not find any probabilistic solution in the first $L$ iterations of the algorithm with probability greater than \[1-\left(\frac{\delta}{\xi(\alpha)}\right)^{\frac{\mu}{\eta}}\Phi(\frac{\alpha\mu}{\eta}, \lceil \log_2\,L\rceil),\]
where the function $\Phi(s,t)$ is given by
\[\Phi(s,t):= \bsis{ll} \fracg{1-2^{(1-s)( t+1)}}{1-2^{1-s}} & \mbox {\rm  if } s\neq 1 \\ \\ t +1 & \mbox{\rm otherwise}\\\esis \]
where $s$ is a strictly positive scalar and $t$ is a non-negative integer.
\end{property}

\ourproof
We notice that $a=0$ implies that, at iteration $k$, the algorithm classifies a candidate solution $\hat{\theta}_k$ as a probabilistic solution only if it satisfies the constraint $g(\hat{\theta}_k,v^{(k)})=0$, $k=1,\ldots,M_k$ where $\{ v^{(1)}, \ldots, v^{(M_k)}\}$ is the randomly obtained validation set $\setV_k$.
Since $E(\theta) \geq \mu$ for all $ \theta\in \Theta$ and $a=0$, the probability of classifying a candidate solution as a probabilistic solution at iteration $k$ is no greater than
\[\left(1-\mu\right)^{M_k}={\rm e}^{M_k \ln  (1-\mu)}  <   {\rm e}^{ -\mu M_k} \le \]
\[\leq  {\rm e}^{-\frac{\mu}{\eta} \ln\,\left( \frac{\xi(\alpha) k^\alpha}{ \delta} \right)} =  \left( \frac{\delta}{\xi(\alpha) k^\alpha} \right)^{\frac{\mu}{\eta}}.\]
Therefore, the probability of providing a probabilistic solution at any of the first $L$ iterations of the algorithm is smaller than
\begin{eqnarray*}
\Sum{k=1}{L} \left( \frac{\delta}{\xi(\alpha) k^\alpha} \right)^{\frac{\mu}{\eta}} & = & \left(\frac{\delta}{\xi(\alpha)}\right)^{\frac{\mu}{\eta}} \Sum{k=1}{L}
\left(\frac{1}{k^\alpha}\right)^\frac{\mu}{\eta}.
\end{eqnarray*}
Taking $s=\frac{\alpha\mu}{\eta}$  and using Property \ref{prop:Riemann} in the Appendix we have
\[
\Sum{k=1}{L}
\left(\frac{1}{k^\alpha}\right)^\frac{\mu}{\eta}= \Sum{k=1}{L} \frac{1}{k^s} \leq \Phi(s,\lceil \log_2\,L\rceil).
\]
We conclude that the probability of not finding any probabilistic solution in the first $L$ iterations of the algorithm is
smaller than
\[ 1-\left(\frac{\delta}{\xi(\alpha)}\right)^{\frac{\mu}{\eta}}\Phi(\frac{\alpha\mu}{\eta}, \lceil \log_2\,L\rceil).\]
\hfill $~$  \qed

We now present an example demonstrating that a strict validation scheme may not be well-suited for a robust design problem.

\noindent
\begin{example}
Suppose that $\Theta=[0,1]$, $\setW=[-0.08,1]$, $\eta=0.1$, $\delta=10^{-4}$ and
\[g(\theta,w)= \bsis{rc} 0 & \mbox {\rm if } \theta\leq w\\
1& \mbox{\rm otherwise.} \esis \]
Suppose also that ${\rm Pr}_{\setW}$ is the uniform distribution. It is clear that $\theta=0$ minimizes the probability of violation and satisfies $\eta = 0.1 > E(0)=\frac{0.08}{1.08}>0.074$. Therefore, we obtain
$E(\theta) \geq 0.074 = \mu \mbox{ for all }\theta\in \Theta.$
Consider now the choice $\alpha=1.1$ and a maximum number of iterations $L$ equal to $10^6$.  We conclude from Property \ref{property:never} that, regardless of the strategy used to obtain candidate solutions, the choice $a=0$ and $\alpha=1.1$ in Corollary \ref{coro:algorithm}, no probabilistic solution with probability greater than $0.98$ is found. The choice $\alpha=2$ leads to a probability greater than $0.99$.  This illustrates that a strict validation scheme is not well suited for this robust design problem.

\hfill $~$  \qed
\end{example}

The next result states that the probabilistic validation scheme presented in this paper achieves, under minor technical assumptions, a solution with probability one in a finite number of iterations.

\begin{property}\label{prop:itworks}
Consider an SPV algorithm with given accuracy parameter $\eta\in(0,1)$, confidence $\delta\in(0,1)$ and level function $m(\cdot)$. Suppose that
\blista \item $\mu(\cdot)$ is a failure function.
\item The cardinality function $M(k,\eta,\delta)$ is given by
 \[\left\lceil \frac{1}{\eta} \left(m(k)+\ln\,\frac{1}{\delta \mu(k)}+\sqrt{2m(k)\ln\,\frac{1}{\delta \mu(k)}}\, \right) \right\rceil. \]
\item There exist an integer $k^*$, scalars $\mu\in(0,1)$ and $p\in(0,1)$ such that at every iteration $k>k^*$ a candidate solution $\hat{\theta}_k$ satisfying  $E(\hat{\theta}_k)\leq \mu <\eta$ is obtained with probability greater
    than $p$.
\item $\lim\limits_{k\to\infty} \frac{1}{m(k)}\ln\,\frac{1}{\delta \mu(k)} = 0.$
\elista
Then, the SPV algorithm achieves with probability one a solution in a finite number of iterations.
\end{property}

\ourproof
Using the assumption  \[\lim\limits_{k\to\infty} \frac{1}{m(k)}\ln\,\frac{1}{\delta \mu(k)} = 0\]
we conclude that
\[\lim\limits_{k\to\infty} \frac{M(k)}{m(k)} = \]
\[\lim\limits_{k\to \infty} \frac{1}{\eta} \left(1+\frac{1}{m(k)}\ln\,\frac{1}{\delta \mu(k)} \right. \left.+\sqrt{2\frac{1}{m(k)}\ln\,\frac{1}{\delta\mu(k)}}\, \right) =\frac{1}{\eta}.\]

Since $\mu< \eta$ and $\frac{m(k)}{M(k)}$ converges to $\eta$, then there exists
$\tilde{k}$ such that
\[\frac{\mu+\eta}{2} \leq \frac{m(k)}{M(k)},\; \mbox{for every } k>\tilde{k}.\]
That is, at each iteration $k$, the SPV algorithm provides candidates solutions $\hat{\theta}_k$ satisfying
\[E(\hat{\theta}_k) \leq \mu \leq \frac{\mu+\eta}{2} \leq \frac{m(k)}{M(k)}\]
for every $k\geq \max\{ k^*,\tilde{k}\}$ with probability greater than $p$. We notice that $\frac{1}{M(k)} \Sum{\ell = 1}{M(k)} g(\hat{\theta}_k,v^{(\ell)}) $
is the empirical mean associated to $g(\hat{\theta}_k,v)$. We recall that the Chernoff inequality (see \cite{Tempo13}) guarantees that the probability of obtaining an empirical mean larger than
$\epsilon=\frac{\eta-\mu}{2}$ from the value $E(\hat{\theta}_k)$ is no larger than ${\rm e}^{-2M(k)\epsilon^2}$.
Notice that
\begin{eqnarray*} E(\hat{\theta}_k) + \epsilon &=& E(\hat{\theta}_k) + \frac{\eta-\mu}{2} \\
&  \leq  &\mu + \frac{\eta-\mu}{2} =  \frac{\mu+\eta}{2} \leq \frac{m(k)}{M(k)}.
 \end{eqnarray*}
 Therefore we have that if $k\geq \max\{ k^*,\tilde{k}\}$ then with probability no smaller than $1-{\rm e}^{-2M(k)\epsilon^2}$ the candidate solution is classified as a probabilistic solution.
 Taking into account that $M(k)$ tends to infinity with $k$, there exists $k_{\epsilon}$ such that $1-{\rm e}^{-2M(k)\epsilon^2}\geq \frac{1}{2}$ for every $k> k_\epsilon$. This means that the probability of classifying a candidate solution as a probabilistic one is no smaller than $\frac{p}{2}$ for every iteration $k>\max\,\{ k^*,\tilde{k},k_{\epsilon}\}$. Since $\frac{p}{2}>0$, we conclude that the algorithm obtains a probabilistic solution with probability one. \hfill $~$  \qed

\section{Numerical example}\label{sec:applications}


The objective of this numerical example is to obtain probabilistic upper and lower bounds of a given time function $y:\setW \to \mathds{R}$ with unknown parameters $A$ and $B$ of the form
\[y(w)=[ A(1+\frac{1}{2}t^2)\sin(7 t+0.5)+B]{\rm e}^{-\frac{3}{2} t},\]
where $w \in \setW$. The uncertainty set $\setW$ is  \[ \setW = \left\{w=[t\quad A\quad B]^T,  t\in[0,1], \;A\in [1,3],\; B\in[1,3]\right\}.\]
For a given order $d$, we define the regressor $\varphi_d:\setW \to \mathds{R}^{d+1}$ as
\[\varphi_d(w)= \varphi_d([t\quad A\quad B]^T ) = \bmat{ccccc} 1 & t & t^2 & \cdot \cdot \cdot & t^d\emat^T .\]
The objective of this example is to find a parameter vector $\theta=[\gamma_d,\lambda_d]^T$, $\gamma_d\in \mathds{R}^{d+1}$ and $\lambda_d\in\mathds{R}^{d+1}$ such that, with probability no smaller than $1-\delta$,
\[ {\rm Pr}_{\setW} \set{w \in \setW}{ |y(w)-\gamma_d^T\varphi_d(w)| \geq \lambda_d^T |\varphi_d(w)| } \leq \eta.\]
The vector $|\varphi_d(w)|$ is obtained from the absolute values of $\varphi_d(w)$. The binary function $g:\Theta\times \setW\to \{0,1\}$, is defined
as
 \[g(\theta,w):=\bsis{rcl} 0 & &\mbox{if } \theta \mbox{ meets design specifications for  } w \\ 1 & & \mbox{otherwise}, \esis \]
where ``design specifications'' means satisfying the following constraint:
\[ |y(w)-\gamma_{d}^T\varphi_d(w)|   \leq  \lambda_{d}^T   |\varphi_d(w)| \]
for uniformly randomly generated samples $w \in \setW$.

A similar problem is addressed in  \cite{Campi08} using the scenario approach. For the numerical computations, we take $\delta=10^{-6}$ and $\eta=0.01$. We address the problem studying the finite families, scenario and SPV approach, and use the explicit sample complexity derived in the previous sections.

\subsection{Finite families approach} \label{subsec:applicationsFinite}

We apply the results of Section \ref{subsec:finite} to determine both the degree $d$ and the parameter vectors $(\gamma_d,\lambda_d)$ that meet the design specification and optimize a given performance index.

In this example a finite family of cardinality $n_C=400$ is considered. In order to compare the finite family approach with the scenario one, we consider no allowed failures (i.e $m=0$). For this choice of parameters ($m=0$, $n_C=400$, $\delta=10^{-6}$ and $\eta=0.01$), the number of samples $N$ required to obtain a solution with the specified probabilistic probabilities is $1981$ (see Theorem \ref{theorem:finite:explicit}). A set $\setD$ of $M=N$ samples is drawn (i.i.d.) from $\setW$. We use these samples to select the optimal parameters $(\tilde{\gamma}_{d}, \tilde{\lambda}_{d})$ corresponding to each of the different regressors $\varphi_d(\cdot)$. Each pair
$(\tilde{\gamma}_{d}, \tilde{\lambda}_{d})$ is obtained minimizing the empirical mean of the absolute value of the approximation error. That is, each pair
$(\tilde{\gamma}_{d}, \tilde{\lambda}_{d})$ is the solution to the optimization problem
\begin{eqnarray*} \;\;\;\;\;\;\min\limits_{\gamma_{d},\lambda_{d}} &\;&\frac{1}{M}\Sum{w\in \setD}{} \lambda_d^T |\varphi_d(w)|    \\
\;\;\;\;\;\; s.t. & \;& |y(w)-\gamma_{d}^T\varphi_d(w)|   \leq  \lambda_{d}^T   |\varphi_d(w)| , \; \forall w\in \setD.
\end{eqnarray*}
We notice that  the obtained parameters do not necessarily satisfy the probabilistic design specifications. In order to resolve this problem, we consider a new set of candidate solutions of the form
$ \Theta=\set{\theta_{d,j}=(\tilde{\gamma}_{d}, {\rm e}^{\left( -0.5+\frac{j}{20} \right)}\tilde{\lambda}_{d})}{
d=1,\ldots,d_{\max}, j=1,\ldots,j_{\max}}.$

This family has cardinality $n_C=d_{\max} j_{\max}$. We take a large factor ${\rm e}^{\left( -0.5+\frac{j}{20} \right)}$ to increase the probability to meet the design specifications. Therefore, choosing a large enough value for $j_{\max}$ leads to a non-empty intersection of $\Theta$ with the set of parameters that meet the design specifications. In this example, we take $j_{\max}=20$ and $d_{\max}=20$, which yields $n_C=400$.

Using the finite family approach, we choose from $\Theta$ the design parameter that optimizes a given performance index.
 We draw from $\setW$ a set $\setV$ of $N$ (i.i.d.) samples and select the pair that minimizes the empirical mean of the absolute value of the approximation error in the validation set $\setV$. That is, we consider the performance index \[\frac{1}{N}\Sum{w\in \setV}{} {\rm e}^{\left( -0.5+\frac{j}{20} \right)}\tilde{\lambda}_d^T |\varphi_d(w)|  \]
 subject to the constraints
 \[|y(w)-\tilde{\gamma}_{d}^T\varphi_d(w)|   \leq  {\rm e}^{\left(-0.5 +\frac{j}{20} \right)}\tilde{\lambda}_{d}^T   |\varphi_d(w)| , \; \forall w\in \setV.\]
We remark that the feasibility of this problem can be guaranteed in two ways. The first one is to choose $j_{\max}$ large enough. The second one is to allow $m$ failures. As previously discussed, in this example we take $j_{\max}=20$ and $m=0$.

As the cardinality $N$ of $\setV$ has been chosen using Theorem \ref{theorem:finite:explicit}, the probability of violation and the probability of failure of the best solution from $\Theta$ are bounded by $\eta$ and $\delta$ respectively.

\begin{figure} [hbtp]
\centering
\includegraphics[width=9cm]{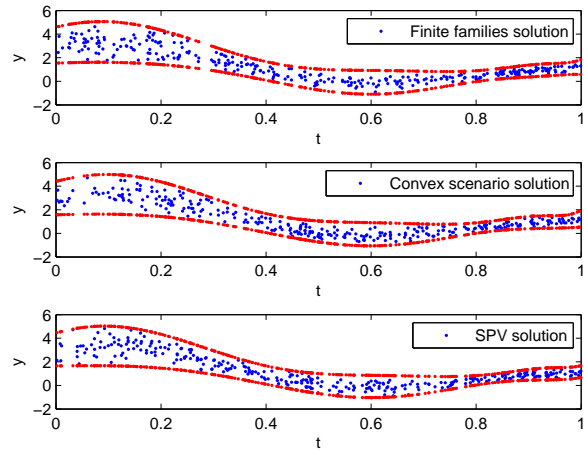}\\
\caption{Initial data set and envelope of the set of solutions.}\label{figuraComparativa}
\end{figure}

The obtained solution corresponds to $d=15$ and $j=11$ and the value for the performance index is $0.9814$.  Figure \ref{figuraComparativa}
shows the approximation for the set $\setV$ and the obtained probabilistic upper and lower bounds for the random function.

Finally, for illustrative purposes, we used a validation set of sample size $N_v=10 N$, obtaining a number of $5$ violations. The empirical violation probability turned out to be $\eta_{\exp} =\frac{5}{10N}=2.5242\cdot 10^{-4}$, while the specification was $\eta=0.01$.

\subsection{Convex scenario approach}

In this case we take advantage of the result of Subsection \ref{subsec:applicationsFinite} and choose $d=15$ as the order of the approximation polynomial. Following the scenario approach we draw a set $\setW_k$ of  $N$ samples (i.i.d) from $\setW$ and solve the convex optimization problem
 \begin{eqnarray*} \;\;\;\;\;\;\min\limits_{\gamma_{d},\lambda_{d}} &\; &\lambda_{d}^{T} E\{|\varphi_d(t)|\}   \\*
 \;\;\;\;\;\; s.t. &\;&  |y(w)-\gamma_{d}^T\varphi_d(w)|   \leq  \lambda_{d}^T   |\varphi_d(w)| , \; \forall w\in \setW_k.
 \end{eqnarray*}
In order to guarantee the design specifications we use Theorem \ref{theorem:convexscenario} to determine the value of $N$. Since the number of decision variables is $2(d+1)=32$, $\eta=0.01$ and $\delta=10^{-6}$, the resulting value for $N$ is $7090$. We notice that the convex scenario approach does not apply directly to the minimization of the empirical mean. This is why one has to resort to the exact computation of the mean of the approximation error $\lambda_{d}^{T} E\{|\varphi_d(t)|\}$, see \cite{Campi08}. Figure (\ref{figuraComparativa}) shows the initial data set generated using the procedure described above, plus the envelope that contains all the polynomials.

For illustrative purposes, we check with a validation set of size $N_v=10 N$. The experimental value $\eta_{\exp}=8.4626\cdot10^{-5}$ is obtained, while the specification was $\eta=0.05$. Using this strategy, $7090$ samples are required, considerably bigger than in the finite families approach. We obtained a performance index of $0.9613$, slightly better than that obtained by the finite families strategy. The advantage of the finite families approach is that, using a smaller number of samples, a similar performance is obtained. This allows us to determine the best order of the polynomial with the further advantage that the exact computation of the mean of the error is not required. Furthermore, the finite family approach does not rely on a convexity assumption.

\subsection{SPV algorithm}

We again take advantage of the result of Subsection \ref{subsec:applicationsFinite} and choose $d=15$ as the order of the approximation polynomial. Following the SPV algorithm approach, we begin setting $\eta=0.01$, confidence $\delta=10^{-6}$, scalars $a= 0.75$, $\alpha=2$ and iteration index $k=1$. The initial $\setW_k$ is a set of $500$ samples drawn from $\setW$ according to probability ${\rm Pr}_{\setW}$.

\blista
\item
A candidate solution $\hat{\theta}_k$ to the problem
 \begin{eqnarray*} \min\limits_{\gamma_{d},\lambda_{d}} &\; &\lambda_{d}^{T} \sum |\varphi_d(t)|\   \\
s.t. &\;&  |y(w)-\gamma_{d}^T\varphi_d(w)|   \leq \lambda_{d}^T   |\varphi_d(w)| , \; \forall w\in \setW_k \\
 \end{eqnarray*}
is obtained.

\item  Set $m_k=\left\lfloor ak \right\rfloor $ and  \[M_k= \left\lceil \frac{1}{\eta} \left(m_k+\ln\frac{\xi(\alpha)k^{\alpha}}{\delta}+\sqrt{2m_k\ln\frac{\xi(\alpha)k^{\alpha}}{\delta}}\, \right) \right\rceil.\]
 \item Obtain validation set $\setV_k =\{v^{(1)},\ldots,v^{(M_k)}\}$ drawing $M_k$ i.i.d. validation samples from $\setW$ according to the probability ${\rm Pr}_{\setW}$.
\item If  $\Sum{\ell=1}{M_k} g(\hat{\theta}_k,v^{(\ell)})\leq m_k$, then $\hat{\theta}_k$ is a probabilistic solution.
\item Exit if the exit condition is satisfied.
\item $k=k+1.$ $\setW_k=\setW_{k-1}\bigcup \setV_{k-1}$. Goto (i).
\elista

Figure (\ref{figuraComparativa}) shows the initial data set generated using the procedure described above, and the envelope that contains all the solution polynomials. Using this strategy, $4163$ samples are required. We obtained a performance index of 0.9406, slightly better than the ones obtained by the other approaches.

The level function in the last step of the algorithm is $m_k=1$, being the empirical probability of failure $\frac{1}{M_k}=\frac{1}{2231}<0.01.$
\begin{remark}
If we set $a=0$ in the algorithm, there are no allowed failures and this coincides with the approach studied in \cite{Oishi07}. In this case, the algorithm did not find a solution for $\eta=0.05$ and $M_k<30000$. This is consistent with the results of Section \ref{sec:Oishi}.
\end{remark}

\begin{table}[hptb]
\begin{center}
\begin{tabular}{| c | c | c |c |}
	\hline
	$\eta$&$N_{\rm finite}$	&	$N_{\rm convex}$	&	$N_{\rm SPV}$	\\
	\hline
	0.1&  398	& 488 & 988 \\
	\hline
	0.05& 794& 849 &1972\\
	\hline
    0.01& 3962 & 7090 & 4163\\
    \hline
    0.005& 7924 & 16078&13652\\
    \hline
    0.001& 39614 & 74062 &41617\\
    \hline
\end{tabular}
\caption{Required sample complexity for different values of $\eta$.}
\end{center}
\end{table}

In Table 1 the results of the three approaches are compared for different values of $\eta$. Note that $N_{\rm{finite}}$, $N_{\rm{convex}}$ and $N_{\rm{SPV}}$ denote the total number of samples required in each of the three proposed strategies. We notice that, for small values of the probability of violation $\eta$, the sample complexity corresponding to the convex scenario is the largest one. On the other hand, as can be observed in Table 2, the performance index obtained with the SPV algorithms is slightly better than the ones corresponding to the other two approaches. We recall that the SPV algorithms do not rely on a convexity or finite cardinality assumptions.

\begin{table}[hptb]
\begin{center}
\begin{tabular}{| c| c | c |c |}
	\hline
	$\eta$&$J_{\rm finite}$	&	$J_{\rm convex}$	&	$J_{\rm SPV}$	\\
	\hline
	0.1&   1.0411	& 0.8803 &  0.8589\\
	\hline
	0.05& 1.0085 & 0.9217 & 0.9597\\
	\hline
    0.01& 0.9841 & 0.9613 & 0.9406\\
    \hline
    0.005& 1.0111 & 1.0447 & 0.9741\\
    \hline
    0.001& 0.9904 & 1.0183 & 0.9828\\
    \hline
\end{tabular}
\caption{Obtained performance index for different values of $\eta$.}
\end{center}
\end{table}

\section{Conclusions}\label{sec:conclusion}

In this paper, we have derived {\it sample complexity} for various analysis and design problems related to uncertain systems.
In particular, we provided new results which guarantee that a binomial distribution is smaller than a given probabilistic confidence.
These results are subsequently exploited for analysis problems to derive the sample complexity of worst-case performance and robust optimization. With regard to design problems, these results
can be used for finite families and for the special case when the design problem can be recast as a robust convex optimization problem.

We also presented a general class of randomized algorithms based on probabilistic validation, denoted as
{\it Sequential Probabilistic Validation} (SPV). We provided a strategy to adjust the cardinality of the validation sets to guarantee that the obtained solutions meet the probabilistic specifications.
The proposed strategy is compared with other existing schemes in the literature. In particular, it has been shown that a strict validation strategy where the design parameters need to satisfy the constraints for all the elements of the validation set might not be appropriate in some situations. We have shown that the proposed approach does not suffer from this limitation because it allows the use of a non-strict validation test.
As it has been shown in this paper, this relaxed scheme allows us to reduce, in some cases dramatically, the number of iterations required by the sequential algorithm. Another advantage of the proposed approach is that it does not rely on the existence of a robust feasible solution. Finally, we remark that this strategy is quite general and it is not based on finite families or convexity assumptions.

\begin{ack}                               
This work was supported by the MCYT-Spain and the European Commission which funded this work under projects DPI2010-21589-C05-01, DPI2013-48243-C2-2-R and FP7-257462. The work of Roberto Tempo was
supported by the European Union Seventh Framework Programme [FP7/2007-2013] under grant agreement n. 257462
HYCON2 Network of Excellence.

\end{ack}

\bibliography{BibNew}

\appendix

\section{Appendix: Auxiliary proofs and properties}\label{sec:app}

{\bf Proof of Corollary \ref{corollary:sqrt}:}
We first notice that if $m=0$,  then  $B(N,\eta,0)=(1-\eta)^N=\ee^{N\ln\, (1-\eta)}\leq \ee^{-\eta N}$.
Therefore, it follows from $\eta N\geq \ln \frac{1}{\delta}$ that $\ee^{-\eta N}\leq \ee^{\ln\, \delta}=\delta$.
This proves the result for $m=0$.

Consider now the case $m>0$. We first prove that
$
h(r):= \sqrt{2(r-1)} - \ln\, \left( r+\sqrt{2(r-1)}
\right)\geq 0$ for all $r\geq 1.$
 Since $h(1)=0$, the inequality
$h(r)\geq 0$ holds if the derivative of $h(r)$ is strictly positive for every $r>1$.
\[\frac{{\rm{d}}}{{\rm d}r}h(r)  = \]
\[\frac{1}{\sqrt{2(r-1)}}-\frac{1}{r+\sqrt{2(r-1)}}\left(1+\frac{1}{\sqrt{2(r-1)}}\, \right)  =\]
 \[ \left(\frac{1}{\sqrt{2(r-1)}}\right)\left(1- \frac{1+\sqrt{2(r-1)}}{r+\sqrt{2(r-1)}}\, \right)  = \]
 \[ \left(\frac{1}{\sqrt{2(r-1)}}\right)\left(\frac{r-1}{r+\sqrt{2(r-1)}}\, \right)\geq 0,\;\; \forall r>1.\]
This proves the inequality $h(r)\geq 0$, for all $r\geq 1$.
Denote now $\hat{a}=r+\sqrt{2(r-1)}$, with $r=1+\frac{1}{m}\ln\,\frac{1}{\delta}$. Clearly $\hat{a}>1$. Therefore, from a direct application of Lemma \ref{property:a:geq:one}, we conclude that it suffices to choose $N$ such that
\[ N\eta \geq \frac{\hat{a}}{\hat{a}-1}\left(\ln\,\frac{1}{\delta}+m\ln\, \hat{a}\right) =\]
  \[\frac{r+\sqrt{2(r-1)}}{r-1+\sqrt{2(r-1)}}\left(r-1+\ln\, (r+\sqrt{2(r-1)}\right)m. \]
Since $h(r)\geq 0$ we conclude that
\[\frac{r-1+\ln\,(r+\sqrt{2(r-1)})}{r-1+\sqrt{2(r-1)}} \leq 1.\]
From this inequality, we finally conclude that inequality $B(N,\eta,m)\leq \delta$ holds if
\[ N\eta \geq (r+\sqrt{2(r-1)})m = m+ \ln\,\frac{1}{\delta}  + \sqrt{2m\ln\,\frac{1}{\delta}}.\]
\hfill $~$  \qed

\begin{property}\label{prop:decreasingBinomial}
For fixed values of $m$ and $N$, $m<N$, the binomial distribution function $B(N,\eta,m)$ is a strictly decreasing function of $\eta\in(0,1)$.
\end{property}

\ourproof
To prove the property, we show that the derivative of $B(N,\eta,m)$ with respect to $\eta$ is negative.
Let us define  the scalars $\varphi_i$, $i=0,\ldots,N$ as follows
\begin{eqnarray}
\varphi_i (\eta)&:= &  \frac{\rm d}{\rm d\eta} \left( \eta^i (1-\eta)^{N-i}\right)\nonumber\\
& = & i\eta^{i-1}(1-\eta)^{N-i}-(N-i)\eta^i(1-\eta)^{N-i-1} \nonumber\\
& = & (i(1-\eta)-(N-i)\eta)\eta^{i-1}(1-\eta)^{N-i-1} \nonumber\\
& = & (i-N\eta)\eta^{i-1}(1-\eta)^{N-i-1} \label{equ:signiNeta}.
\end{eqnarray}
With this definition we have
\begin{eqnarray}
\frac{\rm d}{\rm d \eta} B(N,\eta,m) &=&  \frac{\rm d}{\rm d \eta} \Sum{i=0}{m}\conv{N}{i}\eta^i(1-\eta)^{N-i} \nonumber\\
& = & \Sum{i=0}{m}\conv{N}{i}\varphi_i(\eta).\label{equ:derieta}
\end{eqnarray}
We consider here two cases, $m-N\eta<0$ and $m-N\eta\geq 0$. In the first case we have from equation (\ref{equ:signiNeta}) that $\varphi_i(\eta)<0$, for $i=0,\ldots,m$. This fact, along with equation (\ref{equ:derieta}) implies that the derivative with respect to $\eta$ is negative and therefore the claim of the property is proved for this case.

Consider now the case  $m-N\eta\geq 0$. In this case we have that $\varphi_i(\eta)>0$, for $i>m$. Since $m<N$ we obtain
\begin{eqnarray*}
\frac{\rm d}{\rm d \eta} B(N,\eta,m) &=& \Sum{i=0}{m}\conv{N}{i}\varphi_i(\eta)\\
& < & \Sum{i=0}{N}\conv{N}{i}\varphi_i(\eta) \\
& = & \Sum{i=0}{N}\conv{N}{i} \frac{\rm d}{\rm d \eta} \left( \eta^i (1-\eta)^{N-i}\right) \\
& = & \frac{\rm d}{\rm d \eta} \Sum{i=0}{N}\conv{N}{i}  \eta^i (1-\eta)^{N-i} \\
& = & \frac{\rm d}{\rm d \eta} (\eta +(1-\eta))^N =
\frac{\rm d}{\rm d \eta}(1)^N =0.
\end{eqnarray*}
We notice that in the last step of the proof the identity
\[(x+y)^N = \Sum{i=0}{N}\conv{N}{i} x^iy^{N-i} \]
has been used.
\qed

\begin{property}\label{prop:Riemann}
Suppose that $L$ is a positive integer and that $s$ is a strictly positive scalar. Then,
 $ \Sum{k=1}{L} \frac{1}{k^s} \leq \Phi(s,\lceil \log_2\,L\rceil)$
 where, given $s\geq 0$ and the integer $t\geq 0$,
 \[ \Phi(s,t):= \bsis{ll} \fracg{1-2^{(1-s)( t+1)}}{1-2^{1-s}} & \mbox { if } s\neq 1 \\ \\ t +1 & \mbox{otherwise}.\\\esis\]
\end{property}

\ourproof
Given $L>0$ and $s>0$, define $t:=\lceil \log_2(L)\rceil$ and $S(t):=\Sum{k=1}{2^t} \frac{1}{k^s}$.
Then we have $ \Sum{k=1}{L} \frac{1}{k^s} \leq \Sum{k=1}{2^t} \frac{1}{k^s}=S(t). $
Next we show that $S(t)\leq 1+2^{1-s}S(t-1)$ for every integer $t$ greater than 0.
Since $S(0)=1$ and $S(1)=1+2^{-s}$, the inequality is clearly satisfied for $t=1$. We now prove the inequality for $t$ greater than 1
\begin{eqnarray*}
S(t) & = & \Sum{k=1}{2^t} \frac{1}{k^{s}} = \Sum{k=1}{2^{t-1}} \left[ \frac{1}{(2k)^s} + \frac{1}{(2k-1)^s} \right]\\
& = & 2^{-s} \Sum{k=1}{2^{t-1}} \frac{1}{k^s}+\Sum{k=1}{2^{t-1}} \frac{1}{(2k-1)^s} \\
& \leq & 2^{-s} S(t-1) + 1 + \Sum{k=2}{2^{t-1}} \frac{1}{(2k-2)^s} \\
& = & 2^{-s} S(t-1) + 1 + 2^{-s} \Sum{k=1}{2^{t-1}-1} \frac{1}{k^s} \\
& \leq & 2^{-s} S(t-1) + 1 + 2^{-s} \Sum{k=1}{2^{t-1}} \frac{1}{k^s} \\
&=& 1+2^{1-s}S(t-1).
\end{eqnarray*}
We have therefore proved the inequality $S(t)\leq 1+2^{1-s}S(t-1)$ for every integer $t$ greater than 0. Using this inequality in a recursive way with $S(0)=1$ we obtain $ S(t)\leq \Sum{k=0}{t} 2^{(1-s)k} = \Phi(s,t). $
This proves the result.
\hfill $~$  \qed

\end{document}